\documentclass[useAMS,usenatbib,fleqn]{mn2e}

\usepackage{times}

\usepackage{graphicx}
\usepackage{epsfig}
\usepackage{amssymb,amsmath}
\usepackage{aas_macros}

\title[Inhomogeneous discs and black hole X-ray
binaries]{Inhomogeneous accretion discs and the soft states of \\black hole
  X-ray binaries}
\author[Dexter \& Quataert]{Jason Dexter$^{1}$\thanks{E-mail: 
jdexter@berkeley.edu} and Eliot Quataert$^{1}$\\
$^{1}$Department of Astronomy and Theoretical Astrophysics Center,
University of California, Berkeley, CA 94720-3411, USA\\
}
\begin{document}
\newcommand{\lnsigt}{\ln{(10)}\hspace{2pt}\sigma_T}
\pagerange{\pageref{firstpage}--\pageref{lastpage}} \pubyear{2012}
\maketitle

\label{firstpage}

\begin{abstract}
Observations of black hole binaries (BHBs) have established a rich
phenomenology of X-ray states. The soft states range from
the low variability, accretion disc dominated thermal (TD) state to
the higher variability, non-thermal steep power law state (SPL). The
disc component in all states is typically modeled with standard thin disc accretion
theory. However, this theory is inconsistent with optical/UV
spectral, variability, and gravitational microlensing observations of
active galactic nuclei (AGNs), the supermassive analogs of BHBs. An
inhomogeneous disc (ID) model with large ($\simeq 0.4 \rm$ dex)
temperature fluctuations in each radial annulus can qualitatively
explain all of these AGN observations. The inhomogeneity may be a
consequence of instabilities in radiation-dominated discs, and therefore
may be present in BHBs as well. We show 
that ID models can explain many features of the TD and SPL states of
BHBs. The observed relationships between  spectral hardness, disc
fraction, and rms variability amplitude in BHBs are reproduced with
temperature fluctuations similar to those inferred in AGNs, suggesting
a unified picture of luminous accretion discs across orders of
magnitude in black hole mass. This picture can be tested with 
spectral fitting of ID models, X-ray polarization observations, and
MHD simulations of radiation-dominated discs. If BHB accretion discs
are indeed inhomogeneous, only 
the most disc-dominated states (disc fraction $\gtrsim 0.95$) can be
used to robustly infer black hole spin using current continuum fitting
methods.
\end{abstract}


\section{Introduction}

Spectral and timing observations of black hole X-ray binaries (BHBs)
in the energy range $\sim 1-100$ keV have established characteristic
outburst ``states'' common to many 
sources \citep[e.g.,][hereafter RM06]{remillardmcclintock2006}. The
``thermal'' (TD) state has little variability, and a large fraction of the
integrated flux is well described by a standard thin black hole
accretion disc model \citep[NT,][]{shaksun1973,novthorne}. The
``low/hard'' state is characterized by a hard power law spectrum with
a weak, soft disc component. It is often modeled as a thin disc
truncated well outside of the black hole marginally stable orbit. Inside
of this location, the accretion flow is assumed to be 
geometrically thick and radiatively inefficient
\citep{esinetal1997}. The ``steep power law'' state (SPL) is
characterized by a soft (photon index $\Gamma > 2.4$) power
law extending unbroken to $\lesssim 1$ MeV \citep{groveetal1998}. Its
physical origin remains uncertain. 

This common interpretation of the X-ray states relies on the validity
of the NT model. However, this model is theoretically inconsistent in
the TD and SPL states when the disc extends down to
the marginally stable orbit of the black hole. In these luminous inner
regions, radiation pressure provides the vertical support against
gravity, and the NT model is both thermally
\citep{shakurasunyaev1976} 
and viscously \citep{lightmaneardley1974} unstable. Global MHD
simulations of thin discs \citep[][and references therein]{pennaetal2010,sorathiaetal2012}, which
include the physics of angular momentum transport 
via the magnetorotational instability \citep{mri}, do not include
radiation pressure and therefore cannot test the validity of the NT
model. Local MHD simulations find that small patches of radiation-dominated
discs are thermally stable \citep{hirosestable2009}, but suggest
that the viscous (Lightman-Eardley) instability may operate
\citep{hiroseetal2009}. This instability has also been proposed as a
model for spectral state transitions in GRS
1915+105 \citep{bellonietal1997,neilsenetal2012}.

The NT model is disfavored by optical/UV
observations of active galactic nuclei (AGNs) with similar $L/L_{\rm
  edd}$ to the soft, luminous states of BHBs. The model underproduces the
observed UV emission \citep[][]{zhengetal1997} and requires a
relativistic mechanism to explain 
the simultaneous variability observed at well separated optical
wavelengths \citep[][]{kroliketal1991}. Recent microlensing
observations find that quasar accretion discs are a factor of $\sim 4$
larger than predicted by the NT model \citep[][and references
therein]{jimenezvicenteetal2012}. \citet[][hereafter
DA11]{dexteragol2011} showed that a disc  
with large, local temperature fluctuations ($\simeq 0.4$ dex)
can explain all of these observations. The fluctuations could be
driven by instabilities in radiation pressure dominated discs, in
which case they would operate in BHBs as well. 

In this Letter, we propose a novel interpretation of the soft,
luminous TD to SPL states of BHBs in terms of this inhomogeneous accretion disc
model (ID, summarized in \S \ref{sec:model}). We show how interpreting
ID spectra with the standard NT model can explain the
X-ray spectral and variability properties of soft  
BHBs (\S \ref{sec:states}). In \S\ref{sec:power-law-tail} we assess
the requirements for the ID model to fit observed BHB X-ray
spectra. Inhomogeneity also may have
important implications (\S \ref{sec:impl-cont-fitt}) for the
continuum fitting method for measuring black hole spin 
\citep[CF, ][and references therein]{mcclintocketal2011}. We discuss the
prospects for testing BHB accretion disc inhomogeneity observationally and
theoretically in \S\ref{sec:discussion}.

\section{Inhomogeneous accretion discs}
\label{sec:model}

DA11 proposed a toy model of an inhomogeneous accretion disc. The disc
is taken to be optically thick everywhere so that its flux 
is specified completely by the effective temperature,
$T(r,\phi,t)$. The radial temperature profile is set by assuming that the model
reproduces the NT surface brightness when 
averaged over azimuth and time. Retaining the NT model on
average is motivated theoretically by
conservation laws and numerical simulations \citep{hiroseetal2009},
and is consistent with quasar microlensing observations
\citep{eigenbrodetal2008}. DA11 broke the disc into
independently fluctuating zones spaced evenly in $\log r$ and $\phi$,
with $n$ total zones per octave in radius. Motivated by results from quasar
variability \citep[e.g.][]{kellyetal2009,macleod2010}, each zone
performs a ``damped random walk'' in $\log T$ around the local thin
disc value with amplitude $\sigma_T$ in dex and timescale $t_{\rm
  drw}$. Both of these quantities are
assumed constant in radius, and the results are insensitive to $t_{\rm
  drw}$. The resulting power spectra are broadly consistent with those 
observed in BHBs \citep{kellyetal2011}, but do not include the high
frequency quasi-periodic oscillations often seen in the SPL. 

For a fixed number of zones, the rms variability increases with increasing
$\sigma_T$. We fix $n=1000$ (see \S~\ref{sec:states}), in which case
the rms is well fit by the empirical formula (cf. equation 1 of DA11), 

\begin{equation}
\label{rms}
\rm rms \simeq 0.02 \sqrt{\exp{\left(16 \sigma_T^2\right)}-1}
\end{equation}

\noindent In the limit $n \rightarrow \infty$, this time-dependent model leads
to a stationary flux profile depending only on $r$ and 
$x \equiv \lnsigt$: 

\begin{equation}
\label{eq:1}
  F_\nu (r;\sigma_T) = \frac{2\pi h \nu^3}{f^4 c^2} \int_0^\infty
  \frac{dw}{\sqrt{\pi }x w}
  \frac{e^{-[\ln{w}+x^2]^2/x^2}}{e^{z/w}-1}, 
\end{equation}

\noindent where $F_\nu$ is the flux at frequency $\nu$
and $z \equiv h \nu / k f T$. The effective temperature in each annulus is
integrated over a log-normal distribution, whose width is $\ln{(10)}
\sigma_T/\sqrt{2}$. The local spectrum is assumed to be a
colour-corrected blackbody with $f=1.7$
\citep[e.g.,][]{shimuratakahara1995}. 

\begin{figure}
\begin{center}
\includegraphics[scale=0.75]{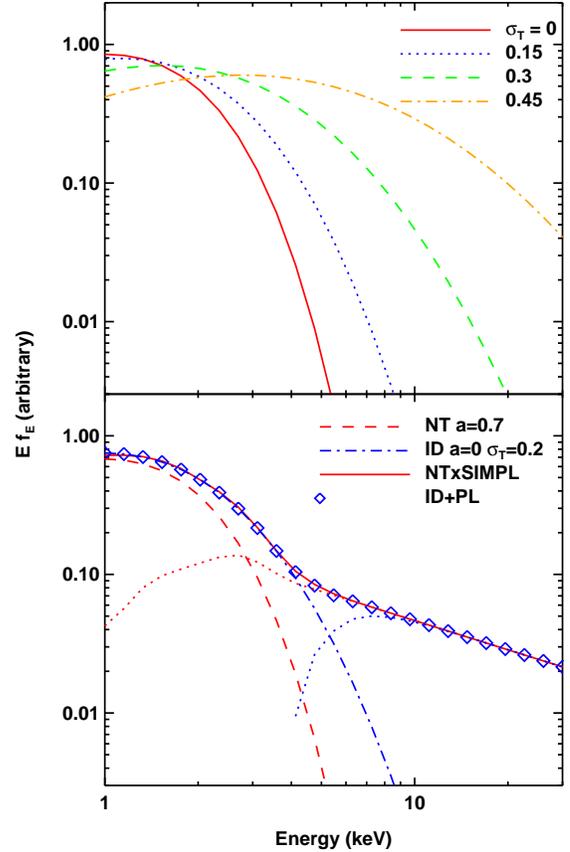}
\caption{\label{comp_plots}Top: X-ray spectra from the inhomogeneous
  disc (ID) model with $a=0.5$. The spectra become harder and broader with
  increasing $\sigma_T$. Bottom: X-ray spectra from the NT (dashed)
  and NT$\otimes$SIMPL (solid) models. A broader ID spectrum (dash-dotted
  line) with a power law tail (open diamonds) can appear nearly
  identical to the NT$\otimes$SIMPL spectrum. The 
  power law components are shown as dotted lines. The
  black hole mass is $10M_\odot$, and the SIMPL parameters are
  $f_{SC}=0.2$, $\Gamma=2.7$.}
\end{center}
\end{figure}

In the following sections, this stationary log-normal model is used
for spectral calculations and the rms variability is estimated from
equation \eqref{rms}. Using the log-normal model is valid for $n
\gtrsim 30$. Except in the brief discussion of X-ray 
polarization in \S\ref{sec:discussion}, all relativistic and viewing geometry effects on the 
observed spectra are ignored. Sample ID spectra are
shown in the top panel of Figure~\ref{comp_plots}.

\begin{figure}
\includegraphics[scale=0.7]{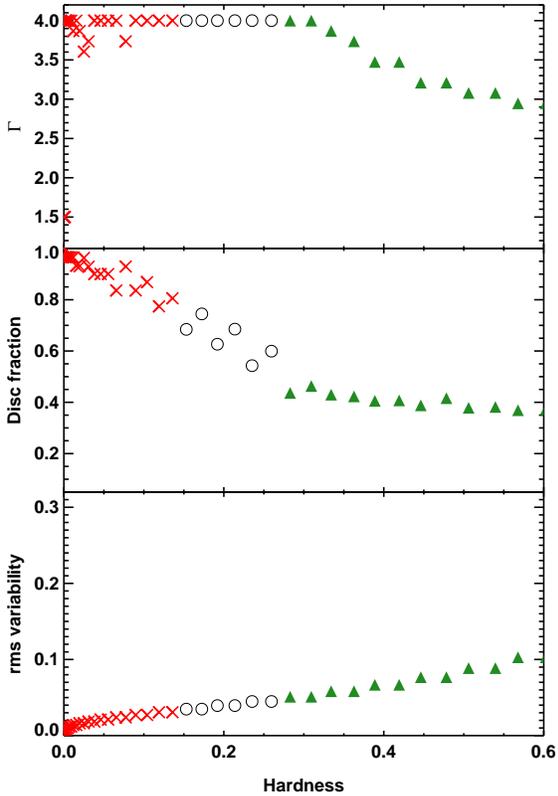}
\caption{\label{states}Inferred spectral and variability
  properties of ID models with
  $a=0.5$. Quantities are defined and plotted as in panels 
  e), f), and g) of Figs. 4-9 of 
  RM06. Each point is the result of fitting an
  ID spectrum with parameters in the range
  $\sigma_T=0.02-0.5$ and $\dot{m}=0.8$, $1.2$ with the NT$\otimes$SIMPL
  model. The points are displayed according to their X-ray spectral state: thermal (red Xs),
  intermediate (open circles), and steep power law (green triangles).}
\end{figure}

\begin{figure}
\includegraphics[scale=0.58]{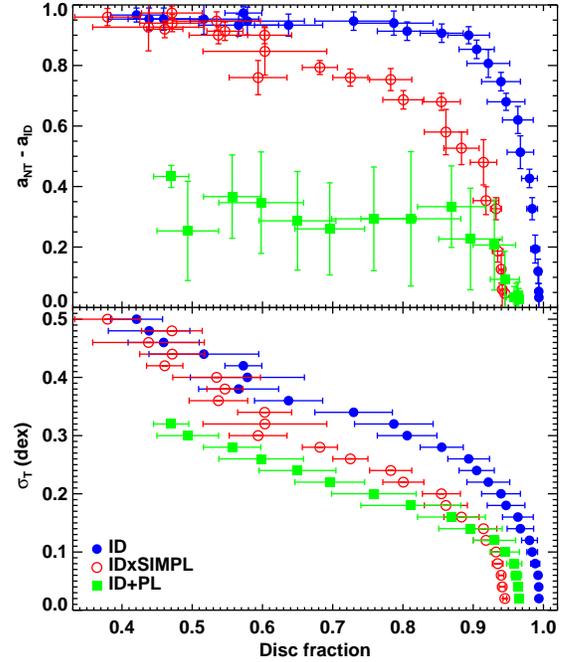}
\caption{\label{deltaa}Top: Black hole spin inferred from fitting 
  $a=0$ ID spectra with the NT$\otimes$SIMPL model as a function
  of disc fraction. If the ID
model correctly describes BHBs, the NT interpretation of the data
can significantly overestimate the true spin for disc fractions
$\lesssim 0.95$. The magnitude of the effect is sensitive to the
properties of the high energy spectral tail. Bottom: $\sigma_T$ vs. disc fraction for the same
fits. The disc fraction rapidly decreases for $0.15 \lesssim \sigma_T
\lesssim 0.35$, where the ID spectra become significantly broader than
their NT counterparts. The ID spectra either have no high energy tail (blue filled circles), a
  power law matched on arbitrarily above 5 keV (green filled squares), or a
  power law tail from SIMPL (red open circles) with $f_{\rm
    SC}=0.03$, $\Gamma=2.5$. The
error bars are the ranges inferred  
from models with varying $\dot{m}$ at fixed $\sigma_T$.}
\end{figure}

\section{Thermal and steep power law states}
\label{sec:states}

RM06 define three states of BHBs in terms of their
spectral and variability characteristics. The hardness ratio, the
ratio of integrated flux from $8.6-18.0$ keV to that from 
$5.0-8.6$ keV, and the rms variability are model-independent. The disc
fraction is the ratio of the integrated disc to total flux between
$2-20$ keV. The disc flux is found by fitting spectra with the NT
model, as well as a power law tail component with a corresponding photon
index $\Gamma$. 

Qualitatively, the ID models reproduce the observed
trends in the TD to SPL states. For small
$\sigma_T \lesssim 0.1$, the spectra are nearly identical to NT
spectra. They are soft, disc-dominated, and exhibit little variability
(TD state). As $\sigma_T$ increases, the intrinsic disc spectrum
becomes broader and harder (Figure \ref{comp_plots}). Fitting with a
thin disc then requires an increasing 
degree of Comptonization, decreasing the disc
fraction. The rms variability also increases with increasing
$\sigma_T$. For large $\sigma_T \gtrsim 0.3$, the ID spectrum extends
to energies $\gtrsim 20$ keV and bears
little resemblance to an NT spectrum, so that the disc fraction
is small. The rms variability has also increased
significantly. This is similar to the observed SPL state. 

We demonstrate this quantitatively by treating the ID spectra as the
data and least-squares fitting them with NT 
models along with the SIMPL prescription for Comptonization
\citep[NT$\otimes$SIMPL,][]{steineretal2009}. With the simulated data and the parameters of
the NT fits, we calculate the required RM06 parameters:
hardness ratio and rms variability from the data (ID spectra), and the disc
fraction and photon index $\Gamma$ from the fits. The parameters of
the ID (NT) spectra are $a$, 
$\dot{m}$, and $\sigma_T$ ($a$, $\dot{m}$, $f_{\rm SC}$, and
$\Gamma$). For a given ``source,'' the black hole spin
is fixed, while $\dot{m}$ and $\sigma_T$ vary. All of the NT
parameters are allowed to vary. The normalized accretion rate is
$\dot{m} \equiv \dot{M} c^2 / L_{\rm edd}$. 

An example is shown in Figure \ref{states}. Each point represents a
value of $\dot{m}$ and $\sigma_T$ from an ID spectrum
with $a=0$, $\dot{m}=0.8$, $1.2$, and $\sigma_T=0.02-0.5$, in $0.02$
intervals. The number of ID zones, $n$, is set by requiring that the
rms variability not exceed $15\%$ in the SPL state for a range of ID
spin values. The inferred value $n=1000$ is
consistent with that required to explain optical/UV observations of
AGNs (Figure 5 of DA11). The maximum
value of $\sigma_T$ that leads to an SPL spectrum is $\simeq 0.5$,
identical to the maximum allowed by AGN observations. Using a smaller
value of $n$ would restrict the range of allowed $\sigma_T$ (rms $<
15\%$), preventing the model from describing the lowest disc
fraction SPL states.

The results in Figure \ref{states} are strikingly similar to the observed behavior of BHBs
(e.g., compare with Figure 4 of RM06). The
relationship between disc fraction and hardness ratio is recovered
with no free parameters. One point on the rms variability vs. hardness
ratio relation is fixed by the choice of $n$, but the predicted
trend is also in agreement with observations. Different 
choices for ID spin, $\sigma_T$, and $\dot{m}$ lead to 
variations of the same trends, and can explain much of the 
diversity observed in soft BHBs. For example, the
presence of distinct tracks in plots of disc fraction vs. hardness
ratio (e.g., Figure 7 of RM06) can be explained as variations in
$\sigma_T$ at different values of $\dot{m}$. The inclusion of changes
in $n$ or possible correlations between $n$, $\sigma_T$, and $\dot{m}$
could cause additional scatter in Figure \ref{states}.

\section{Power law tail models}
\label{sec:power-law-tail}

In the previous Section, the ID spectra were used as data without any
power law spectral tail. This reduces the number of free
parameters, but leads to steeper inferred photon indices than
observed. The origin of this high energy tail emission is uncertain
\citep[e.g.,][]{doneetal2007}. It can 
be modeled as a power law, often extending over all X-ray energies. However,
at energies near the transition between disc and power law components,
emission typically attributed to the power law could be due to a
broader disc spectrum like that produced by the ID models. 

The high energy
tail can also be modeled with physical \citep{titarchuk1994,esinetal1997} or
empirical \citep{steineretal2009} Comptonization models. In both
cases, the disc spectrum (or its Wien tail) is used as the seed
photon spectrum, implicitly assuming that the Comptonizing
corona extends uniformly over the disc. This picture is strongly
disfavored in AGNs by X-ray microlensing observations, where the
corona is found to be much more compact than 
the optical/UV disc \citep{chartasetal2009,daietal2010}. Steep radial
emissivity profiles found from modeling X-ray 
reflection spectra in BHBs and AGNs
\citep{miniuttifabian2004,schmolletal2009} also favor a compact
corona. There is no theoretical reason to expect large differences in
coronal structure between AGNs and BHBs. All models for the high
energy tail, then, introduce considerable uncertainty into BHB
spectral fitting.

This uncertainty may allow the ID model of the SPL state to fit the
data as well as the significantly Comptonized NT one. An example is
shown in the bottom panel of Figure 
\ref{comp_plots}. The same spectrum (solid line) is produced by a significantly
Comptonized NT$\otimes$SIMPL, and a strongly inhomogeneous disc 
supplemented by a power law with a low energy cutoff at $5$ keV
(diamonds). This power law 
could mimic Comptonization from a compact corona. Recent radiative
transfer calculations from relativistic MHD simulations have also found
that emission from the plunging region, neglected in the NT model, can
lead to a power law spectral tail \citep{zhudavisetal2012}. Whatever
the emission mechanism or geometry, the ID model for BHBs 
still requires that the normalization of the high energy tail
increases with decreasing disc fraction (increasing $\sigma_T$). ID
spectra alone  cannot explain the strong, non-thermal emission
extending to $1$ MeV in some SPL states.

\section{Implications for black hole spin measurements via continuum
  fitting}
\label{sec:impl-cont-fitt}

One promising method for measuring black hole spin is to fit disc-dominated 
BHB spectra with the NT$\otimes$SIMPL model. The hardness (or characteristic temperature) of the disc
spectrum depends on its inner edge, which is assumed to coincide with the
marginally stable orbit of the black hole. This allows a measurement
of the spin, provided that the mass, inclination, and distance are
accurately determined. This technique has been used to estimate the
spins of several BHBs \citep[e.g.,][]{mcclintocketal2011}. 

In the ID model, the characteristic temperature
increases with both increasing spin and 
$\sigma_T$. Therefore, if BHB accretion discs are inhomogeneous, the
use of the NT model could introduce systematic errors in CF spin measurements. From the
fitting procedure used in \S\ref{sec:states}, the magnitude of these errors can be estimated as a
function of the inferred disc fraction. The results are shown in the
top panel of Figure \ref{deltaa} for an ID model with $a=0$. The range of
$\sigma_T$ used is the same as in Figure 
\ref{states}, and is consistent with that favored by optical/UV AGN
observations ($0.35 < \sigma_T < 0.50$, DA11). 

The inferred spin using the NT model is always larger than the ``true'' ID spin, since the
spectral peak moves to higher frequencies with increasing
$\sigma_T$. Quantitatively, the spin differences depend on the
high energy tail model used. This is because the spin and the SIMPL
scattering efficiency $f_{\rm SC}$ both harden the spectrum. Without
any high energy tail (blue filled circles), $f_{\rm SC}$ is small for
$\sigma_T < 0.15$ and instead the NT spin increases to fit the
intrinsically harder ID spectra. A power law matched on smoothly to the ID 
spectrum above $5$ keV (green filled squares) leads to a higher
power law normalization with increasing $\sigma_T$. In that case, $f_{\rm SC}$
must increase to fit the tail, and much smaller changes in spin can explain
the harder ID spectra. The disc fraction
decreases sharply for $\sigma_T \gtrsim 0.15$ as the ID spectra become
broad, requiring significant Comptonization of the NT spectra in the
fits (bottom panel of Figure \ref{deltaa}). This leads to the sharp 
transition in inferred spin below disc fraction $\simeq 0.93-0.97$, where
the transition value depends on the high energy tail model. 

\section{Discussion}
\label{sec:discussion}

X-ray spectral and timing observations of BHBs have established a set
of states common to many objects. The physical
origin of the SPL state and the cause of state transitions remain
poorly understood. We have shown that the 
spectral and rms variability properties of soft, luminous BHBs (TD to
SPL states) 
can be explained by varying
one parameter: the amplitude of local temperature
fluctuations in an inhomogeneous accretion disc. Comparable
temperature fluctuations in quasar accretion discs can resolve a
number of important observational puzzles in AGNs (DA11). The ID model then
provides a unified picture of luminous black hole accretion 
discs across many orders of magnitude in black hole mass. 

The most likely physical processes for driving large temperature
fluctuations are radiation pressure 
dominated disc instabilities. Local MHD simulations are consistent with
the thin disc prediction for the relationship between surface density
and stress (or temperature) when time-averaged 
\citep{hiroseetal2009}. The simplest interpretation is that the disc
is viscously unstable. If so, it would likely produce
temperature fluctuations qualitatively similar to those in the ID
model. This can be tested with radially-extended radiation MHD
simulations. Alternatively, large fluctuations could also occur in 
magnetically-dominated discs \citep[e.g.,][]{begelmanpringle2007},
where the disc radiation energy content is de-coupled from its stability. 

The ID model is insufficient to fit the X-ray spectra observed in both
AGNs and BHBs. The broad ID spectrum produces some, but not all, of
the emission typically attributed to the high energy tail, and some prescription for the
remaining emission is still required. This could be from
Comptonization in a compact corona or from hot gas inside the 
disc marginally stable orbit \citep{zhudavisetal2012}. As with other 
models, the high energy tail normalization must increase with decreasing
disc fraction in order to explain SPL spectra. Quantitative fitting of
the ID spectral model to observations will be carried out in future
work, with various prescriptions for the high energy tail. Spectral models could
use the colour-corrected  blackbody used here, or could incorporate the results of
sophisticated accretion disc atmosphere  calculations
\citep[BHSPEC,][]{davisetal2005,davisetal2006}.

If BHBs are indeed inhomogeneous, spin measurements from current
CF methods using the NT model are potentially subject to large
systematic uncertainties (spin differences $\Delta a \sim 0.3-0.9$ for
ID spin $a=0$, Figure \ref{deltaa}). These can be much
larger than current statistical uncertainties and systematic errors 
from ignoring emission from the plunging region
\citep[$0.2-0.3$,][]{kulkarnietal2011,nobleetal2011}, except in the
most disc-dominated states (disc fraction $\gtrsim 0.95$ or $\sigma_T
\lesssim 0.15$). Therefore, if BHB discs are significantly
inhomogeneous, spin measurements 
using CF methods should be restricted to the most
disc-dominated states.

The ID model also generically
predicts that the inferred spin using NT models should increase with decreasing disc
fraction. \citet{steinersilver2009,steineretal2010,steineretal2011} have found
that inferred spins in a few sources do not change significantly over
a wide range of $L/L_{\rm edd}$ or $f_{\rm SC}$. Since $f_{\rm SC}$ is a proxy for disc 
fraction, this latter conclusion is in apparent conflict with our
Figure~\ref{deltaa} \citep[particularly Figure 1
of][]{steinersilver2009}. However, the magnitude of the inferred spin
differences when fitting NT models to ID spectra depends sensitively on the high energy tail model
assumed. Further, the inferred spin from fitting ID models is
nearly constant outside of a critical range in disc fraction. Observations of IDs 
entirely above (below) this range in disc fraction could find a constant, correct
(incorrect) spin value, although there is no evidence for sharp
changes in spin from CF at disc fractions $\simeq 0.95$. Fitting X-ray spectra with the ID model will allow 
estimates of $a$ and $\sigma_T$ in a variety of
states, and test the model. Requiring the spin to be independent of
spectral state may constrain $\sigma_T$ and the degree of
inhomogeneity in BHB accretion discs. 

X-ray polarization measurements may provide another test for
inhomogeneous accretion discs in BHBs. Assuming a semi-infinite
scattering-dominated atmosphere \citep{chandrasekhar1950}, we have
calculated polarized images and spectra via ray tracing using the codes
\textsc{geokerr} \citep{dexteragol2009} and \textsc{grtrans}
\citep{dexter2011}. We account for all relativistic effects, including
the rotation of the polarization vector
\citep{connorspiranstark1980,agolphd}. With these assumptions, the 
time-averaged ID polarization is unchanged from the
NT model: $1-5\%$ peak polarization, increasing strongly from face-on to
edge-on viewing, and also increasing at lower black hole
spin \citep{schnittmankrolik2009}. X-ray polarization may then provide an
additional constraint on the BH spin, independent of the location of
the spectral peak. In addition, the polarization angle and degree vary
by factors of a few on the fluctuation timescale. Both the degree of 
polarization and its time-variability increase with increasing
$\sigma_T$. These estimates ignore returning radiation, as well as coronal
\citep{schnittmankrolik2010} and plunging region 
\citep{agolkrolik2000} emission. These effects could
cause additional differences between the polarization
signatures of the ID and NT models.

The ID model used here assumes an optically thick disc. The NT 
effective optical depth is \citep[e.g.,][]{abramowiczfragile2011}: 

\begin{equation}
\tau_{\rm NT} \sim 1 \left(\frac{L}{L_{\rm edd}}\right)^{-2}
\left(\frac{R}{10 M}\right)^{93/32},
\end{equation}

\noindent with $G=c=1$ and ignoring the effects of general
relativity and of the inner disc edge on the optical depth. If the
disc is inhomogeneous, portions will become optically thin. Assuming
independent temperature and density fluctuations, we
estimate the fraction of the disc with $\tau_{\rm ID} < 1$ by integrating the
log-normal distribution, $f(w)$, for $\tau_{\rm ID} (w) = w^{2.3} \tau_{\rm NT}
< 1$. For $\sigma_T < 0.1$, $\gtrsim 99\%$
of the disc is optically thick where the bulk of the
luminosity is produced for $L/L_{\rm edd} \le 1$. At larger
$\sigma_T$, the inner disc will become optically 
thin ($\gtrsim 50\%$ by area at the peak emission radius for $\sigma_T >
0.35$ and $L/L_{\rm edd}=1$). If this emission is similar to that from
the plunging region \citep{zhudavisetal2012}, it could lead to the observed 
increasing high energy tail emission with decreasing disc fraction. 
Comptonization from hot electrons in optically thin regions may also
effectively truncate the disc spectrum as inferred in the SPL state
\citep[][]{donekubota2006}.  

\section*{acknowledgements}
JD thanks E. Agol, O. Blaes, J. McClintock, J. Steiner, and J. Tomsick for useful
discussions. EQ is supported in part by the David and Lucile Packard
Foundation. 

\footnotesize{
\bibliographystyle{mn2e}

}
\label{lastpage}

\end{document}